\documentclass[conference]{IEEEtran}

\IEEEoverridecommandlockouts

\usepackage{cite}
\usepackage{amsmath,amssymb,amsfonts}
\usepackage{algorithmic}
\usepackage{graphicx}
\usepackage{textcomp}
\usepackage{xr}
\usepackage[table]{xcolor}

\def\BibTeX{{\rm B\kern-.05em{\sc i\kern-.025em b}\kern-.08em
		T\kern-.1667em\lower.7ex\hbox{E}\kern-.125emX}}

\usepackage{tikz}
\usetikzlibrary{backgrounds}
\usetikzlibrary{spy}
\usetikzlibrary{calc,arrows.meta}
\usepackage{pgfplots}
\usetikzlibrary{plotmarks}
\usetikzlibrary{arrows.meta}
\usetikzlibrary{positioning,fit}
\usetikzlibrary{intersections}
\usetikzlibrary{patterns}
\usetikzlibrary{decorations.shapes}

\usepgfplotslibrary{patchplots}
\usepgfplotslibrary{colormaps}

\usepackage{cancel}

\usepackage{pdfpages} 

\usepackage{multirow}
\usepackage{makecell}

\usepackage{algorithmic}
\usepackage[ruled,vlined]{algorithm2e}
\SetKwRepeat{Do}{do}{while}

\newcommand{\exportFigures}{true}
\newcommand{\exportFiguresAsPNG}{true}

\ifthenelse{\equal{\exportFigures}{true}}
{
	\usepgfplotslibrary{external}
	\tikzexternalize[prefix=compiled_tikz_figures/,optimize command away=\includepdf]
	\ifthenelse{\equal{\exportFiguresAsPNG}{true}}
	{
		\tikzset
		{   png export/.style={
				external/system call={
					pdflatex \tikzexternalcheckshellescape -halt-on-error --extra-mem-top=10000000 -interaction=batchmode -jobname "\image" "\texsource" && pdftops -eps "\image.pdf" && convert -density 700 -transparent white "\image.pdf" "\image.png"
		}}}
		\tikzset{png export}
	}
	{}
}
{}

\usepackage{url}

\usepackage{bm}
\usepackage[caption=false,font=footnotesize]{subfig}
\usepackage[normalem]{ulem}

\usepackage{mathtools, cuted}

\usepackage{acronym}

\usepackage{booktabs}
		
\usepackage{pdfpages}

\usepackage[latin1]{inputenc}
\usepackage{tikz}
\usetikzlibrary{shapes,arrows}
\usetikzlibrary{arrows.meta}
\usetikzlibrary{positioning}
\usetikzlibrary{spy,backgrounds}
\usetikzlibrary{math}

\usepackage{ellipsis}

\usetikzlibrary{calc}
\usetikzlibrary{decorations.pathreplacing,decorations.markings,shapes.geometric}
\usetikzlibrary{decorations.pathmorphing}
\usetikzlibrary{fit}
\usetikzlibrary{pgfplots.groupplots}

\usetikzlibrary{calc,arrows.meta}

\usetikzlibrary{backgrounds} 

\definecolor{green(pigment)}{rgb}{0.0, 0.65, 0.31}
\definecolor{frenchblue}{rgb}{0.0, 0.45, 0.73} 
\definecolor{mediumcandyapplered}{rgb}{0.89, 0.02, 0.17}

\usepackage[most]{tcolorbox}
\tcbuselibrary{breakable}
\tcbset{every box/.style={enhanced,breakable}}
\tcbset{colframe=black,colback=red!10,enhanced,breakable,sharp corners}

\usepackage{enumitem} 

\usepackage{notation}

\usepackage{adjustbox}

\definecolor{alex}{RGB}{51,183,150}
\definecolor{erik}{RGB}{235,134,52}
\definecolor{Vicky}{RGB}{235,134,52}

\newcommand{\ticked}{$\text{\rlap{$\checkmark$}}\square$}
\newcommand{\unticked}{{$\square$}}
\newcommand{\tick}[1]{\ifthenelse{#1=1}{\ticked}{\unticked}}

\newcommand{\rmv}{\hspace*{-.3mm}}

\newcommand{\vm}[1]{\ensuremath{\bm{#1}}} 

\renewcommand{\Re}[1]{\ensuremath{\text{Re}\!\left[#1\right]}}
\renewcommand{\Im}[1]{\ensuremath{\text{Im}\!\left[#1\right]}}

\newcommand{\norm}[2]{\ensuremath{\lVert #1 \rVert^{#2}}}

\newcommand{\minus}{\rmv - \rmv}

\newcommand{\s}{\hspace*{0.5pt}}

\providecommand{\norm}[1]{\lVert#1\rVert}

\newcommand{\nn}{\nonumber}

\mathchardef\Re="023C
\mathchardef\Im="023D

\newlength{\figureheight}
\newlength{\figurewidth}
\graphicspath{{./figures/}}

\allowdisplaybreaks

\definecolor{mycolor01}{rgb}{0.00000,0.00000,1.00000}
\definecolor{mycolor02}{rgb}{0.133,0.545,0.133}
\definecolor{mycolor03}{rgb}{0.50000,0.00000,0.50000}
\definecolor{mycolor05}{rgb}{1.00000,0.83984,0.00000}
\definecolor{mycolor04}{rgb}{0.92969,0.50781,0.92969}
\definecolor{mycolor06}{rgb}{1.00000,0.64453,0.00000}
\definecolor{mycolor07}{rgb}{0.50000,0.50000,0.50000}
\definecolor{mycolor08}{rgb}{1.00000,0.00000,0.00000}
\definecolor{mycolor09}{rgb}{0.2510 ,0.8784, 0.8157}
\definecolor{mycolor10}{rgb}{0.54297,0.00000,0.00000}
\definecolor{mycolor11}{rgb}{0.6445, 0.1641,0.1641}
\definecolor{mycolor12}{rgb}{1, 0, 1}

\pgfdeclarelayer{back1}
\pgfdeclarelayer{lay1}
\pgfdeclarelayer{back2}
\pgfdeclarelayer{lay2}
\pgfsetlayers{back2,lay2,back1,main,lay1}

\makeatletter

\tikzset{
	nomorepostactions/.code={\let\tikz@postactions=\pgfutil@empty},
	decmark/.style 2 args={decoration={markings,
			mark= between positions 0 and 1 step (1/6)*\pgfdecoratedpathlength with{%
				\tikzset{#2,every mark}\tikz@options
				\pgftransformresetnontranslations
				\pgfuseplotmark{#1}%
			},  
		},
		postaction={decorate},
		/pgfplots/legend image post style={
			mark=#1, mark options={#2}, every path/.append style={nomorepostactions}
		},
	},
	markbeginend/.style 2 args={decoration={markings,
			mark= between positions 0 and 1 step (1)*\pgfdecoratedpathlength with{%
				\tikzset{#2,every mark}\tikz@options
				\pgfuseplotmark{#1}%
			},  
		},
		postaction={decorate},
		/pgfplots/legend image post style={
			mark=#1,mark options={#2},every path/.append style={nomorepostactions}
		},
	},
	markend/.style 2 args={decoration={markings,
			mark= at position \pgfdecoratedpathlength with{%
				\tikzset{#2,every mark}\tikz@options
				\pgfuseplotmark{#1}%
			},  
		},
		postaction={decorate},
		/pgfplots/legend image post style={
			mark=#1,mark options={#2},every path/.append style={nomorepostactions}
		},
	},
	posmark/.style 2 args={decoration={markings,
			mark= at position #2 with{%
				\tikzset{solid,every mark}\tikz@options
				\pgftransformresetnontranslations
				\pgfuseplotmark{#1}%
			},  
		},
		postaction={decorate},
		/pgfplots/legend image post style={
			mark=#1,mark options={solid},every path/.append style={nomorepostactions}
		},
	},
}

\makeatother

\pgfplotsset{
	resultStyle1/.style={mark=none, line width=0.5pt, mycolor01, decmark={oplus}{solid}},
	resultStyle2/.style={mark=none, line width=0.5pt, mycolor02, decmark={triangle}{solid}},
resultStyle3/.style={mark=none ,line width=0.5pt, mycolor03, decmark={+}{solid}},
resultStyle4/.style={mark=none, line width=0.5pt, mycolor06, decmark={star}{solid}},
resultStyle5/.style={mark=none, line width=0.5pt, mycolor08, decmark={o}{solid}},
resultStyle6/.style={mark=none, line width=0.5pt, mycolor05, decmark={square}{solid}}, 
resultStyle7/.style={mark=none, line width=0.5pt, mycolor09, decmark={diamond}{solid}}, 
resultStyle8/.style={mark=none, line width=0.5pt, mycolor11, decmark={otimes}{solid}}, 
resultStyle9/.style={mark=none, line width=0.5pt, mycolor12, decmark={x}{solid}}, 
resultStyleBase/.style={mark=none, line width=0.5pt,}, 
compareStyle1/.style={mark=none, line width=0.5pt, mycolor01},
compareStyle2/.style={mark=none, line width=0.5pt, mycolor02},
compareStyle3/.style={mark=none ,line width=0.5pt, mycolor03},
compareStyle4/.style={mark=none, line width=0.5pt, mycolor06},
compareStyle5/.style={mark=none, line width=0.5pt, mycolor08},
compareStyle6/.style={mark=none, line width=0.5pt, mycolor05}, 
compareStyle7/.style={mark=none, line width=0.5pt, mycolor09}, 
compareStyle8/.style={mark=none, line width=0.5pt, mycolor11}, 
compareStyle9/.style={mark=none, line width=0.5pt, mycolor12}, 
}

\pgfplotsset{
compat=newest,
simple style group/.style={
label style={font=\scriptsize},
legend style={font=\scriptsize},
tick label style={font=\scriptsize},
nodes near coords style={font=\scriptsize},
title style={font=\scriptsize},
scale only axis,
grid style={dotted},
mark options={solid}, 
},
simple style/.style={
label style={font=\scriptsize},
legend style={font=\scriptsize},
tick label style={font=\scriptsize},
nodes near coords style={font=\scriptsize},
title style={font=\scriptsize},
width=\figurewidth,
height=\figureheight,
at={(0\figurewidth,0\figureheight)},
scale only axis,
grid style={dotted},
mark options={solid}, 
},
base style/.style={
label style={font=\scriptsize},
legend style={font=\scriptsize},
tick label style={font=\scriptsize},
nodes near coords style={font=\scriptsize},
title style={font=\scriptsize},
width=\figurewidth,
height=\figureheight,
at={(0\figurewidth,0\figureheight)},
scale only axis,
cycle list={
	{mark=none, line width=0.5pt, mycolor01, solid},
	{mark=none, line width=0.5pt, mycolor02, dash dot},
	{mark=none ,line width=0.5pt, mycolor03, densely dashed},
	{mark=none, line width=0.5pt, mycolor04, dash dot dot},
	{mark=x   , line width=0.5pt, mycolor05},
	{mark=.   , line width=0.7pt, mycolor06}, 
	{mark=square,only marks, mark size = 0.8pt, mycolor07,
		mark options = {line width = 0.4pt}},
	{mark=x,     only marks, mark size = 1.3pt, mycolor08,
		mark options = {line width = 0.4pt}},
	{mark=o,     only marks, mark size = 0.8pt, mycolor09,
		mark options = {line width = 0.4pt}},
	{mark=o, mycolor10},
},
grid style={dotted},
xmajorgrids,
ymajorgrids,
mark options={solid}, 
},
base style group/.style={
	label style={font=\scriptsize},
	legend style={font=\scriptsize},
	tick label style={font=\scriptsize},
	nodes near coords style={font=\scriptsize},
	title style={font=\scriptsize},
	scale only axis,
	grid style={dotted},
	xmajorgrids,
	ymajorgrids,
	mark options={solid}, 
},
std graph style new/.style={
xlabel style={yshift=1mm},
ylabel style={yshift=-1.5mm},
yticklabel style={xshift=1mm},
},
color lines style/.style={
cycle list={
	{mark=none, mycolor01, decmark={oplus}{solid} },
	{mark=none, mycolor02, decmark={+}{solid} }, 
	{mark=none, mycolor03, decmark={triangle}{solid} }, 
	{mark=none, mycolor04, decmark={star}{solid} }, 
	{mark=none, mycolor05, decmark={o}{solid} },
	{mark=none, mycolor06, decmark={square}{solid} },
},
},
meas graph style/.style={
xlabel style={yshift=1mm},
ylabel style={yshift=-1mm},
xmajorgrids,
ymajorgrids,
mark repeat = 1,
mark phase = 0,
cycle list={
	{color=black, only marks, mark=*, mark size=0.5pt, mark options={solid, black}},
	{color=red, only marks, mark=*, mark size=0.1pt, line width=0.25pt},
},
ylabel={},
}, 
ci graph style/.style={
xlabel style={yshift=1mm},
ylabel style={yshift=-1.5mm},
yticklabel style={xshift=1mm},
mark repeat = 1,
mark phase = 0,
ymin=1e-3,
ymax=100,
ytick = {100, 50, 10, 1, 0.1, 0.01, 1e-3, 1e-4},
yticklabels = {$0$, $50$, $90$, $99$, $99.9$, $99.99$, $99.999$, $99.9999$},
y dir=reverse,
},     
bp coeff style/.style={
scale only axis=true,
width=0.225*.9\linewidth,
height=0.225*.9\linewidth,
scale only axis,
xmin=-4.000,
xmax=4.000,
xlabel={$\ell${\color{white}$\aod$}},
ticklabel style={font=\footnotesize},
ymin=0.000, ymax=0.9,
ylabel={$c_\ell$},
xlabel style={font=\footnotesize},
ylabel style={font=\footnotesize},
major tick length=2pt
},
bp graph style/.style={        
scale only axis=true,
width=0.35*1.1\linewidth,
height=0.225*.9\linewidth,
scale only axis,
xmin=-3.14, xmax=3.14,
xlabel={$\aod${\color{white}$\ell$}},
ticklabel style={font=\footnotesize},
xtick={-3.14,-1.57,0.0,1.57,3.14},
xticklabels={$-\pi$,$-\tfrac{\pi}{2}$,$0$,$\tfrac{\pi}{2}$,$\pi$},
ymin=0.000, ymax=3,
ylabel={Beampattern},
xlabel style={font=\footnotesize}, ylabel style={font=\footnotesize},
major tick length=2pt
},
peb graph style/.style={        
width=0.66\linewidth,
scale only axis,
point meta min=-2.583,
point meta max=-0.300,
axis on top,
xmin=0.000,
xmax=12.000,
xlabel={x in meter},
y dir=reverse,
ymin=0.000,
ymax=8.000,
ylabel={y in meter},
ytick={7.0,6.0,...,0.0},
xtick={0.0,1.0,...,12.0},
yticklabels={$1$,$2$,$3$,$4$,$5$,$6$,$7$,$8$},
xlabel style={font=\scriptsize,yshift=0.125cm},
ylabel style={font=\scriptsize,yshift=-0.125cm},
ticklabel style={font=\scriptsize},
unit vector ratio*=1 1 1,
yticklabel pos=left,
major tick length=2pt,
colormap={mymap}{[1pt] rgb(0pt)=(1,1,1); rgb(1pt)=(0.858903,0.984776,0.839302); rgb(2pt)=(0.777958,0.94143,0.649487); rgb(3pt)=(0.755504,0.864264,0.463393); rgb(4pt)=(0.777509,0.754439,0.310168); rgb(5pt)=(0.820314,0.619497,0.21003); rgb(6pt)=(0.854796,0.471879,0.170327); rgb(7pt)=(0.851327,0.326629,0.183322); rgb(8pt)=(0.784671,0.198575,0.225774); rgb(9pt)=(0.637629,0.0993149,0.259577); rgb(10pt)=(0.400067,0.0343393,0.229819); rgb(11pt)=(0,0,0)},
colorbar style={ylabel={Position Error Bound in centimeter (logscale)}, ytick={-0.4,-0.82,...,-2.92}, yticklabels={$39.8$, $15.1$, $5.8$, $2.2$, $0.8$, $0.3$},ylabel style={yshift=0.5mm,font=\scriptsize,scale=0.8},width=2.0mm,xshift=-4.25mm,ticklabel style={font=\scriptsize},major tick length=0pt}, 
colormap access=piecewise constant
},
peb ellipses/.style={color=white, line width=0.4pt, forget plot}
}

\tikzset{naming/.style={align=center,font=\small}}
\tikzset{antenna/.style={insert path={-- coordinate (ant#1) ++(0,0.25) -- +(135:0.25) + (0,0) -- +(45:0.25)}}}
\tikzset{station/.style={naming,draw,shape=dart,shape border rotate=90, minimum width=10mm, minimum height=10mm,outer sep=0pt,inner sep=3pt}}
\tikzset{mobile/.style={naming,draw,shape=rectangle,minimum width=12mm,minimum height=6mm, outer sep=0pt,inner sep=3pt}}
\tikzset{radiation/.style={{decorate,decoration={expanding waves,angle=90,segment length=4pt}}}}

\tikzset{
  pobl/.style={
    inner sep=0pt, outer sep=0pt, fill=#1,
  },
  pobl gron/.style n args={2}{
    pobl=#1, rounded corners=#2,
  },
  pics/person/.style n args={3}{
    code={
      \node (-corff) [pobl=#1, minimum width=.25*#2, minimum height=.375*#2, rotate=#3, pic actions] {};
      \node (-pen) [minimum width=.3*#2, circle, pobl=#1, outer sep=.01*#2, anchor=south, rotate=#3, pic actions] at (-corff.north) {};
      \node (-coes dde) [pobl gron={#1}{1pt}, anchor=north west, minimum width=.12125*#2, minimum height=.25*#2, rotate=#3, pic actions] at (-corff.south west) {};
      \node [pobl=#1, anchor=north, minimum width=.12125*#2, minimum height=.15*#2, rotate=#3, pic actions] at (-coes dde.north) {};
      \node (-coes chwith) [pobl gron={#1}{1pt}, anchor=north east, minimum width=.12125*#2, minimum height=.25*#2, rotate=#3, pic actions] at (-corff.south east) {};
      \node [pobl=#1, anchor=north, minimum width=.12125*#2, minimum height=.15*#2, rotate=#3, pic actions] at (-coes chwith.north) {};
      \node (-braich dde) [pobl gron={#1}{.75pt}, minimum width=.075*#2, minimum height=.325*#2, outer sep=.0064*#2, anchor=north west, rotate=#3, pic actions] at (-corff.north east)  {};
      \node [pobl=#1, minimum width=.05*#2, minimum height=.2*#2, outer sep=.0064*#2, anchor=north west, rotate=#3, pic actions] at (-corff.north east) {};
      \node (-braich chwith) [pobl gron={#1}{.75pt}, minimum width=.075*#2, minimum height=.325*#2, outer sep=.0064*#2, anchor=north east, rotate=#3, pic actions] at (-corff.north west) {};
      \node [pobl=#1, minimum width=.0375*#2, minimum height=.2*#2, outer sep=.0064*#2, anchor=north east, rotate=#3, pic actions] at (-corff.north west) {};
      \node (-fit person) [fit={(-pen.north) (-braich dde.east) (-coes chwith.south) (-braich chwith.west)}] {};
    },
  },
  pics/SBS/.style={code={
      \begin{scope}[local bounding box=#1]
      \fill [pic actions/.try] (-1,0) -- (-1/2,3) -- (1/2, 3) -- (1,0) -- cycle;
      \fill [pic actions/.try] (-1/16,2) rectangle (1/16,4);
      \fill [pic actions/.try] (0,4) circle [radius=1/4];
      \foreach \i in {-1,1}
        \fill [shift=(90:4), xscale=\i]
          \foreach \r in {1,3/2,2}{
            (-45:\r) arc (-45:45:\r) -- (45:\r-1/10)
            arc(45:-45:\r-1/10) -- cycle
          };
       \end{scope}
  }},
}

\IEEEoverridecommandlockouts \IEEEpubid{\makebox[\columnwidth]{979-8-3503-8544-1/24/\$31.00~\copyright{}2024 IEEE \hfill} \hspace{\columnsep}\makebox[\columnwidth]{ }}

\begin{document}


\title{Fusion of Active and Passive Measurements for Robust and Scalable Positioning}

\author{Hong Zhu$^{1,2}$, Alexander Venus$^{1,2}$, Erik Leitinger$^{1,2}$, Stefan Tertinek$^{3}$,
 and  Klaus Witrisal$^{1,2}$
\thanks{The financial support by the Christian Doppler Research Association, the Austrian Federal Ministry for Digital and Economic Affairs and the National Foundation for Research, Technology and Development is gratefully acknowledged.}

\\
\small{{$^1$Graz University of Technology, Austria}, {$^3$NXP Semiconductors, Austria},
}\\
\small{{$^2$Christian Doppler Laboratory for Location-aware Electronic Systems}}\\
}


\maketitle
\frenchspacing

\renewcommand{\baselinestretch}{0.98}\small\normalsize 

\begin{abstract}
This paper addresses the challenge of achieving reliable and robust positioning of a mobile agent, such as a radio device carried by a person, in scenarios where direct \ac{los} links are obstructed or unavailable.
The human body is considered as an extended object that scatters, attenuates and blocks the radio signals. 
We propose a novel particle-based \ac{spa} that fuses active measurements between the agent and anchors with passive measurements from pairs of anchors reflected off the body. 
We first formulate radio signal models for both active and passive measurements. Then, a joint  tracking algorithm that utilizes both active and passive measurements is developed for the extended object. The algorithm exploits the \ac{pda} for multiple object-related measurements. 
The results demonstrate superior accuracy during and after the \ac{olos} situation, outperforming conventional methods that solely rely on active measurements. 
The proposed joint estimation approach significantly enhances the localization robustness via radio sensing.

\end{abstract}

\acresetall 


\IEEEpeerreviewmaketitle



\section{Introduction}\label{sec:introduction}

In the realm of wireless communication and location-based services, radio localization stands as a paramount endeavor. 
Among the various technologies being explored, \ac{uwb}  radio is a particularly attractive option for radio sensing applications. UWB radio offers superior temporal resolution, enabling the observation of multipath components (MPCs) from the environment or scattering paths reflected from objects\cite{Gedschold2023,Zetik2007}.
Besides, \ac{uwb} is a wireless communication technology that has been developed to supply localization and sensing in an optimal way. 
This opens up a wide range of possibilities for various functionalities and applications, particularly in the field of mobility and transportation\cite{WitrisalSPM2016Copy,Mendrzik2019,WangShenTWC2020}. Safety-critical  applications, such as autonomous driving\cite{Karlsson2017}, medical services\cite{KoEMBMag2010}, and keyless access systems\cite{Kalyanaraman2020}, greatly benefit from the capabilities offered by radio sensing and communications. 

\begin{figure}[t]
	\centering
    \includegraphics{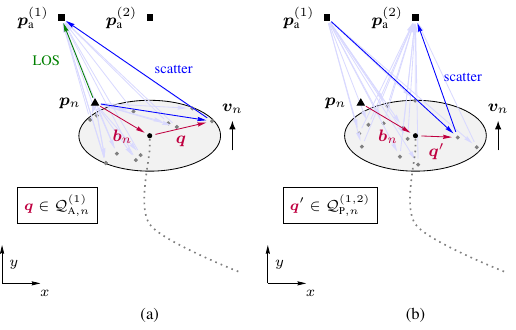}
	\caption{Generic environment with a moving agent and two anchors at time $n$ with exemplary signal propagation paths for (a) an active signal from the agent to anchor 1 and (b) a passive signal between anchor 1 and 2. }
	\label{fig:genericModel}
	\vspace{-6mm}
\end{figure}
 
The human body has indispensable effect on the robustness of the positioning algorithms, as it can cause complete blockage of \ac{los} path to anchors, as well as dispersive and attenuated effects in the radio signals \cite{WildingEUCAP2022}.
The human body can be considered as an extended object. To resolve problems related to \ac{eot}, ellipses\cite{Koch2008,Feldmann2008,Feldmann2011,Schuster2015,Zhang2021}, rectangles\cite{Granstroem2014,Granstroem2011}, and star convex shapes\cite{Baum2014,Hirscher2016,Kumru2021} can be used to model objects. The extended object can generate a diverse range of measurements from spatially dispersed reflection points.
The challenging aspect of the EOT problem is fusing data from multiple object-related measurements at each time step\cite{Meyer2019, MeyerTSP2021, Wielandner2023}. 
\Ac{pda} is a Bayesian method commonly applied in target tracking to handle measurement origin uncertainty\cite{BarShalomTCS2009, Leitinger2023}. 
It can make use of the extracted measurement information, such as delays and amplitudes, to update the state estimate of the object in the presence of clutter\cite{VenusTWC2023, Venus2023}. 
Recently, the \ac{spa} based on factor graphs has been investigated for the tracking of extended objects by exploiting probabilistic data association. 
Note that the proposed SPA has low computational complexity since it scales linearly in the number of measurements. 

In this work, we consider both the measurements between the mobile agent and anchors, referred to as {\textit{active measurements}}, as well as the measurements among anchors reflected off the human body, referred to as \textit{passive measurements}.
We first derive the UWB off-body radio signal models, describing the human body as an extended object. 
Additionally, based on a factor graph designed for joint active and passive estimation, we develop a particle-based \ac{spa} for robust sequential tracking of the extended object, fusing multiple object-related measurements originating from active and passive radar measurements.


\section{Radio Signal Model}\label{sec:signal_model}

At each time step $n$, a mobile agent at position $\bm{p}_n$ transmits a signal, and each anchor $j \rmv\rmv \in \rmv\rmv \{1,\s... \s,J\}$ at position $\bm{p}_{\text{a}}^{(j)} = [p_{\text{ax}}^{(j)} \; p_{\text{ay}}^{(j)}]^\text{T}$ acts as a receiver, capturing active measurements. Synchronously, pairs of anchors $(j,j')$ exchange signals, capturing passive measurements from the human body. Note that the anchors act as receivers and transmitters. The human body is modeled as an extended object (EO) that is rigidly coupled to the mobile agent. The bias between the mobile agent and the body center is denoted as $\bm{b}_n$. 
An example is shown in Fig.~\ref{fig:genericModel}. We refer to the mobile agent that is coupled with the extended object as the extended agent.
For extended-object measurements, we assume the scatters are generated uniformly on the human body, and the human body is regarded as a scattering volume.
We describe the scattering volume for active and passive measurements as $\vm{Q}_{\text{A},n}^{(j)}$ and $\vm{Q}_{\text{P},n}^{(j,j')}$ for received anchor $j$ at time $n$, respectively.
Each point-source scatter from the volume $\vm{Q}_{\text{A},n}^{(j)}$ and $\vm{Q}_{\text{P},n}^{(j,j')}$ is denoted by its position $\vm{q}$ and $\vm{q'}$, respectively\cite{Schubert2013, Wilding2020}. 

\subsection{Active Radio Signal}
At each time step $n$, a radio signal is transmitted from the mobile agent and received at the $j$th anchor (see Fig.~\ref{fig:genericModel}). The complex baseband signal received at $j$th anchor is modeled as
\begin{equation}
	r_{\text{A}, n}^{(j)}(t) =\alpha_{n}^{(j)}s(t-\tau_{n}^{(j)})+\sum_{\vm{q} \in \vm{Q}_{\text{A},n}^{(j)}} \alpha_{\vm{q},n}^{(j)}\beta_{\vm{q}}s(t-\tau_{\vm{q},n}^{(j)})+ w_n^{(j)}(t)
	\label{equ:rx_active}
\end{equation}
where $\alpha_{n}^{(j)}$ and $\tau_{n}^{(j)}$ are the complex amplitude and delay of the LOS component from active measurements. 
The received signal of anchor $j$ also consists of the scatter components originating from the scattering volume $\vm{Q}_{\text{A},n}^{(j)}$.  
The complex amplitude and delay of the scatter component are denoted as $\alpha_{\vm{q},n}^{(j)}$ and $\tau_{\vm{q},n}^{(j)}$, and $\alpha_{\vm{q},n}^{(j)} = \alpha_{\text{A}} / \|(\bm{p}_n+\bm{b}_n+\vm{q}) - \vm{p}_\text{a}^{(j)}\|$, for $\vm{q} \in \vm{Q}_{\text{A},n}^{(j)}$, with $\alpha_{\text{A}}$ describing the amplitude of the transmitted signal from the mobile agent, and $\beta_{\vm{q}}$ is the relative dampening variable for active measurements. 
The second term $w_n^{(j)}(t)$ accounts for measurement noise modeled as \ac{awgn} with double-sided power spectral density $N_0/2$. The propagation delay of the LOS path and scatter path\footnote{Note that here we ignore the distance from the mobile agent to scatter when we calculate the propagation delay of scatter component of active case.} is given by 
\begin{equation}
	\tau_{n}^{(j)} = \|\bm{p}_n - \bm{p}_\text{a}^{(j)}\| /c
	\label{equ:rx_active_LOS}
\end{equation}
\begin{equation}
	\tau_{\vm{q},{n}}^{(j)} = \|(\bm{p}_n+\bm{b}_n+\vm{q}) - \bm{p}_\text{a}^{(j)}\| /c
	\label{equ:rx_active_scatter}
\end{equation}
respectively, where $\vm{q} \in \vm{Q}_{\text{A},n}^{(j)}$ and $c$ is the speed of light. 
\subsection{Passive Radio Signal}
At each time step $n$, a radio signal is transmitted from $j'$th anchor and received at $j$th anchor. The complex baseband signal received at $j$th anchor is modeled as
\begin{equation}
	r_{\text{P}, n}^{(j,j')}(t) = \sum_{\vm{q'} \in \vm{Q}_{\text{P},n}^{(j,j')}}\alpha_{\vm{q'},{n}}^{(j,j')}\beta_{\vm{q}'}s(t-\tau_{\vm{q'},{n}}^{(j,j')}) + w_{n}^{(j,j')}(t)
	\label{equ:rx_passive}
\end{equation}
where $ \alpha_{\vm{q'},{n}}^{(j,j')}$ and $\tau_{\vm{q'},{n}}^{(j,j')}$ are the complex amplitude and delay of the scatter component in $\vm{Q}_{\text{P},n}^{(j,j')}$, and $ \alpha_{\vm{q'},{n}}^{(j,j')} = \alpha_{\text{P}}^{(j')}/(\|(\bm{p}_n+\bm{b}_n+\vm{q}') - \vm{p}_\text{a}^{(j')}\| \|(\bm{p}_n+\bm{b}_n+\vm{q}') - \vm{p}_\text{a}^{(j)}\|)$, for $\vm{q'} \in \vm{Q}_{\text{P},n}^{(j,j')}$, with $\alpha_{\text{P}}^{(j')}$ describing the amplitude of the transmitted signal from anchor $j'$, 
and $\beta_{\vm{q}'}$ is the dampening variable for passive measurement. 
The propagation delay of the passive measurement is given by 
\begin{equation}
	\tau_{\vm{q}',{n}}^{(j,j')} = (\|(\bm{p}_n+\bm{b}_n+\vm{q}') - \vm{p}_\text{a}^{(j')}\| + \|(\bm{p}_n+\bm{b}_n+\vm{q}') - \vm{p}_\text{a}^{(j)}\|)/c
	\label{equ:CIR_passive_delay}
\end{equation}
where $\vm{q'} \in \vm{Q}_{\text{P},n}^{(j,j')}$.
\subsection{Signal Parameter Estimation}\label{eq:CEDA}
At time $n$, measurements which are extracted from the active radio signal are called active measurements $\V{z}_{\text{A},n}$, while measurements which are extracted from the passive radio signal are called passive measurements $\V{z}_{\text{P},n}$.
We define the vectors $\bm{z}_{\text{A},n} = [{\bm{z}_{\text{A},n}^{(1)}}^\text{T} \cdots {\bm{z}_{\text{A},n}^{(J)}}^{\text{T}}]^\text{T}$ and $\bm{z}_{\text{P},n} = [{\bm{z}_{\text{P},n}^{(1,1)}}^{\text{T}} \cdots {\bm{z}_{\text{P},n}^{(J,J)}}^{\text{T}}]^\text{T}$ for the measurement vectors per time $n$. 
For each received anchor $j$, we define $\V{z}^{(j)}_{\text{A},n}= [\V{z}^{(j)}_{\text{A},n,1},\dots\,, \V{z}^{(j)}_{\text{A},n, M_{\text{A}, n}^{(j)}}]$, with $M_{\text{A},n}^{(j)}$ being the number of active measurements. Each active measurement $\V{z}^{(j)}_{\text{A},n,l}= [z_{\text{A},\text{d},n,l}^{(j)} ~ z_{\text{A},\text{u},n,l}^{(j)}]^\text{T}$, $l \in  \{1,\,\dots\,M_{\text{A},n}^{(j)}\}$ contains a distance measurement $z_{\text{A},\text{d},n,l}^{(j)} \in [0, d_\text{max}]$, where $d_\text{max}$ is the maximum measurement distance, and a normalized amplitude measurement $z_{\text{A}, \text{u},n,l}^{(j)} \in [\gamma, \infty)$, where $\gamma$ is the detection threshold.
For the passive case, we define $\V{z}^{(j,j')}_{\text{P},n}= [\V{z}^{(j,j')}_{\text{P},n,1},\dots\,, \V{z}^{(j,j')}_{\text{P},n, M_{\text{P}, n}^{(j,j')}}]$ with $M_{\text{P},n}^{(j,j')}$ being the number of passive measurements. Each passive measurement $\V{z}^{(j,j')}_{\text{P},n,l}= [z_{\text{P},\text{d},n,l}^{(j,j')} ~ z_{\text{P},\text{u},n,l}^{(j,j')}]^\text{T}$, $l \in  \{1,\,\dots\,M_{\text{P},n}^{(j,j')}\}$ contains a distance measurement $z_{\text{P},\text{d},n,l}^{(j,j')} \in [0, d_\text{max}]$ and a normalized amplitude measurement $z_{\text{P}, \text{u},n,l}^{(j,j')} \in [\gamma, \infty)$.

 
\section{Joint Active and Passive Tracking Method}\label{sec:system_model}
 

In this section, we introduce a Bayesian factor graph-based \ac{spa} that fuses joint active and passive measurements. Since the human body is modeled as an extended object, the proposed algorithm is based on the multiple-measurement-to-object data association. 
\subsection{System Model}
At time $n$, the extended agent state is described by a kinematic state and an extent state. The kinematic state  $\bm{x}_n = [\bm{p}_n ^\text{T}\; \bm{v}_n^\text{T}\; \bm{b}_n^\text{T}]^\text{T}$ consists of the mobile agent's position $\bm{p}_n = [p_{\text{x}\s n}\; p_{\text{y}\s n}]^\text{T}$, the velocity $\bm{v}_n = [v_{\text{x}\s n}\; v_{\text{y}\s n}]^\text{T}$, and the bias between the mobile agent and the body center $\bm{b}_n = [b_{\text{x}\s n}\; b_{\text{y}\s n}]^\text{T}$. The shape of the extended object is approximated by an ellipse. The extent state is modeled as $\bm{X}_n = \bm{A}_{n}\bm{E}\bm{A}_{n}^\text{T}$, 
while $\bm{E} \in \mathbb{R}^{2 \times 2}$ is a symmetric, positive semidefinite matrix that describes the 2-D ellipse\footnote{At present, we assume the body shape is not changed over time, thus the extent state is an constant matrix varying with body rotation.}. 
It is further assumed that the square root of the  eigenvalues of $\bm{E}$ is proportional to the object's semi-axis \cite{MeyerTSP2021,Hoher2022}.
The rotation matrix $\bm{A}_n = [\cos(\theta_n), -\sin(\theta_n); \sin(\theta_n), \cos(\theta_n)]$ is formulated with $\theta_n$ describing the body orientation. The body orientation is determined from the velocity states as $\theta_n = \text{atan}(\frac{v_{\text{y}\s n}}{v_{\text{x}\s n}})$.

\subsection{Measurement Model}
\subsubsection{LOS Measurement Model}
The LOS \ac{lhf} of an individual active measurement is given by
\begin{equation} 
	f_{\text{LOS}}(\V{z}^{(j)}_{\text{A},n,l}|\bm{x}_n) = f_{\text{N}}(z_{\text{A},\text{d},n,l}^{(j)}; h_\text{LOS}(\bm{x}_n,\bm{p}_{\text{a}}^{(j)}), \sigma_{\text{d}}^{2} (z_{\text{A},\text{u},n,l}^{(j)})) 
	\label{equ: measLikelihood_active_LOS}
\end{equation}
where $f_{\text{N}}(x;\mu,\sigma^2)$ is the Gaussian PDF, with mean $h_\text{LOS}(\bm{x}_n,\bm{p}_{\text{a}}^{(j)}) = \|\vm{p}_n - \vm{p}_\text{a}^{(j)}\|$ being the LOS measurement function and variance $\sigma_{\text{d}}^{2} (z^{(j)}_{\text{A},\text{u},n,l})$. The variance is determined from the Fisher information given by
$ \sigma_{\text{d}}^{2} (z^{(j)}_{\text{A},\text{u},n,l}) =   c^2 / ( 8\,  \pi^2 \, \beta_\text{bw}^2 \, (z^{(j)}_{\text{A},\text{u},n,l})^2)$, where $\beta_\text{bw}$ is the root mean squared bandwidth \cite{WitrisalJWCOML2016,LeitingerJSAC2015}.
\subsubsection{Scattering Measurement Model}
The scattering \ac{lhf} of individual active and passive measurements are described as
\begin{align}
   &f(\V{z}^{(j)}_{\text{A},n,l}|\bm{x}_n, \bm{\zeta}_{\text{A},n,l}^{(j)}) \nn \\
   &\triangleq f_{\text{N}}(z_{\text{A},\text{d},n,l}^{(j)}; h_\text{A}(\bm{x}_n, \bm{\zeta}_{\text{A},n,l}^{(j)}, \bm{p}_{\text{a}}^{(j)}), \sigma_{\text{d}}^{2} (z_{\text{A},\text{u},n,l}^{(j)}))
   \label{equ:active_scatter_model}
\end{align}
and 
\begin{align}
	&f(\V{z}^{(j,j')}_{\text{P},n,l}|\bm{x}_n, \bm{\zeta}_{\text{P},n,l}^{(j,j')}) \nn \\
	&\triangleq f_{\text{N}}(z_{\text{P},\text{d},n,l}^{(j,j')}; h_\text{P}(\bm{x}_n, \bm{\zeta}_{\text{P},n,l}^{(j,j')}, \bm{p}_{\text{a}}^{(j')},\bm{p}_{\text{a}}^{(j)}), \sigma_{\text{d}}^{2} (z_{\text{P},\text{u},n,l}^{(j,j')}))
	\label{equ:passive_scatter_model}
\end{align}
respectively, where $h_\text{A}(\bm{x}_n,  \bm{\zeta}_{\text{A},n,l}^{(j)},\bm{p}_{\text{a}}^{(j)}) = \|(\vm{p}_n +\bm{b}_n + \bm{\zeta}_{\text{A},n,l}^{(j)})- \vm{p}_\text{a}^{(j)}\|$  and $h_\text{P}(\bm{x}_n,\bm{\zeta}_{\text{P},n,l}^{(j,j')}, \bm{p}_{\text{a}}^{(j')}, \bm{p}_{\text{a}}^{(j)}) = \|(\vm{p}_n +\bm{b}_n+\bm{\zeta}_{\text{P},n,l}^{(j,j')})- \vm{p}_\text{a}^{(j')}\|+ \|(\vm{p}_n +\bm{b}_n +\bm{\zeta}_{\text{P},n,l}^{(j,j')})- \vm{p}_\text{a}^{(j)}\|$  are the nonlinear measurement functions of active scattering measurements and passive scattering measurements, respectively.
Furthermore, we assume that $\bm{\zeta}_{\text{A},n,l}^{(j)}$ and $\bm{\zeta}_{\text{P},n,l}^{(j,j')}$ are the relative positions of scatters with respect to $\bm{p}_n$ that generate $\V{z}^{(j)}_{\text{A},n,l}$ and $\V{z}^{(j,j')}_{\text{P},n,l}$. 
The extent state $\bm{X}_n$ is used to define the covariance matrix of a Gaussian PDF \footnote{It is shown in \cite{Feldmann2011} that for an elliptically shaped object the uniform distribution can be approximated by a Gaussian distribution.} that models the scattering distribution due to the geometric shape as $f(\bm{\zeta}_{\text{A},n,l}^{(j)}|\bm{X}_n) \,\triangleq f_{\text{N}}(\bm{\zeta}_{\text{A}, n,l}^{(j)}; \bm{0}, \bm{X}_n)$ and $f(\bm{\zeta}_{\text{P},n,l}^{(j,j')}|\bm{X}_n)\,\triangleq f_{\text{N}}(\bm{\zeta}_{\text{P},n,l}^{(j,j')}; \bm{0}, \bm{X}_n)$.

The scattering measurement likelihood conditioned on $\bm{x}_n$ and $\bm{X}_n$ is a convolution of the noise distribution and the scattering distribution \cite{MeyerTSP2021}. 
The \ac{lhf} conditioned on $\bm{x}_n$ and $\bm{X}_n$ of an individual scattering measurement can be obtained by integrating out the scattering variables.
Hence, the active scattering \ac{lhf} can be obtained as
\begin{align}
	&f_{\text{AS}}(\V{z}^{(j)}_{\text{A}, n,l}|\bm{x}_n, \bm{X}_n) \nn \\
	&=\int f(\V{z}^{(j)}_{\text{A},n,l}|\bm{x}_n, \bm{\zeta}_{\text{A},n,l}^{(j)})f(\bm{\zeta}_{\text{A},n,l}^{(j)}|\bm{X}_n) d\bm{\zeta}_{\text{A},n,l}^{(j)} \nn\\
	& = f_{\text{N}}(z_{\text{A},\text{d},n,l}^{(j)}; \bar{h}_\text{A}(\bm{x}_n,\bm{p}_{\text{a}}^{(j)}), \sigma_{\text{d}}^{2} (z_{\text{A},\text{u},n,l}^{(j)})+l_{\text{A},n}^{(j)})
	\label{equ: measLikelihood_active_scattering}
\end{align}
where $\bar{h}_\text{A}(\bm{x}_n, \bm{p}_{\text{a}}^{(j)}) = \|(\vm{p}_n +\bm{b}_n)- \vm{p}_\text{a}^{(j)}\|$.
We use the unscented transformation (UT)\cite{Wan2000} for nonlinear transform  of the scatter distribution $f(\bm{\zeta}_{\text{A},n,l}^{(j)}|\bm{X}_n)$ from position domain to delay domain. 
$l_{\text{A},n}^{(j)}$ is the variance of the 
sigma points transformed by the nonlinear function $h_\text{A}(\cdot)$.
The passive scattering \ac{lhf} can be obtained as
\begin{align}
	&f_{\text{PS}}(\V{z}^{(j,j')}_{\text{P},n,l}|\bm{x}_n, \bm{X}_n) \nn\\
	&=\int f(\V{z}^{(j,j')}_{\text{P},n,l}|\bm{x}_n, \bm{\zeta}_{\text{P},n,l}^{(j,j')})f(\bm{\zeta}_{\text{P},n,l}^{(j,j')}|\bm{X}_n) d\bm{\zeta}_{\text{P},n,l}^{(j,j')} \nn \\
	& = f_{\text{N}}(z_{\text{P}, \text{d},n,l}^{(j,j')}; \bar{h}_\text{P}(\bm{x}_n,\bm{p}_{\text{a}}^{(j')},\bm{p}_{\text{a}}^{(j)}), \sigma_{\text{d}}^{2} (z_{\text{P},\text{u},n,l}^{(j,j')})+l_{\text{P},n}^{(j,j')})
	\label{equ: measLikelihood_passive}
\end{align}
where $\bar{h}_\text{P}(\bm{x}_n, \bm{p}_{\text{a}}^{(j')}, \bm{p}_{\text{a}}^{(j)}) = \|(\vm{p}_n +\bm{b}_n)- \vm{p}_\text{a}^{(j')}\|+ \|(\vm{p}_n +\bm{b}_n)- \vm{p}_\text{a}^{(j)}\|$. Similar to the active case, $l_{\text{P},n}^{(j,j')}$ is the variance of the sigma points transformed by the nonlinear function $h_\text{P}(\cdot)$.
\subsubsection{Active and Passive Measurement Model}
The active measurement model including the LOS path and the scattering paths is given as
\begin{align}
	&f_\text{A}(\V{z}^{(j)}_{\text{A},n,l}|\bm{x}_n,\bm{X}_n) \nn\\
	&=f_{\text{LOS}}(\V{z}^{(j)}_{\text{A},n,l}|\bm{x}_n)+ 
	f_{\text{AS}}(\V{z}^{(j)}_{\text{A}, n,l}|\bm{x}_n,\bm{X}_n)
	\label{equ: measLikelihood_active}
\end{align}
and the passive measurement model including only  the scattering paths is given as
\begin{equation}
	f_\text{P}(\V{z}^{(j,j')}_{\text{P},n,l}|\bm{x}_n,\bm{X}_n) 
	=f_{\text{PS}}(\V{z}^{(j,j')}_{\text{P}, n,l}|\bm{x}_n,\bm{X}_n)
	\label{equ: measLikelihood_passive}
\end{equation}

\subsection{Data Association Uncertainty}
For each anchor $j$, the measurements $\V{z}_{\text{A},n}^{(j)}$ and $\V{z}_{\text{P},n}^{(j,j')}$ are subject to data association uncertainty. It is not known which measurement is associated with the \ac{los} path, which measurements originate from the extended object, or if a measurement did not originate from the \ac{los} path and the object, i.e.  clutter. 
The association variables $a_{\text{A},n,l}^{(j)} \in \{0,1\}$ and $a_{\text{P},n,l}^{(j,j')} \in \{0,1\}$  specify whether a single measurement is object-related, which is denoted by $1$, or not, which is denoted by $0$.
Furthermore, let $\V{a}^{(j)}_{\text{A},n}= [{a}^{(j)}_{\text{A},n,1},\dots\,, {a}^{(j)}_{\text{A},n, M_{\text{A}, n}^{(j)}}]$ and $\V{a}^{(j,j')}_{\text{P},n}= [{a}^{(j,j')}_{\text{P},n,1},\dots\,, {a}^{(j,j')}_{\text{P},n, M_{\text{P}, n}^{(j,j')}}]$, as well as $\bm{a}_{\text{A},n} = [{\bm{a}_{\text{A},n}^{(1)}}^\text{T}\; \cdots {\bm{a}_{\text{A},n}^{(J)}}^{\text{T}}]^\text{T}$ and $\bm{a}_{\text{P},n} = [{\bm{a}_{\text{P},n}^{(1,1)}}^{\text{T}}\; \cdots {\bm{a}_{\text{P},n}^{(J,J)}}^{\text{T}}]^\text{T}$. 

The number of object-related measurements is Poisson distributed with mean $\mu_{\text{m}}$. The number of clutter measurement is also Poisson distributed with mean $\mu_{\text{c}}$, and clutter measurements are independent and distributed according to the uniform distribution $f_c(\V{z}_{\text{A},n,l}^{(j)})$ and $f_c(\V{z}_{\text{P},n,l}^{(j,j')})$. 
Conditioned on $\bm{x}_n$ and $\bm{X}_n$, the object-related measurements are independent of the other measurements.
The pseudo-likelihood function for active and passive cases can be represented as \cite{Koch2008}
\begin{align}
	&{g}_{\text{A}}(\V{z}_{\text{A},n,l}^{(j)}|\bm{x}_n, \bm{X}_n, a_{\text{A},n,l}^{(j)})\nn\\
	&\hspace*{15mm}= \begin{cases}
		\frac{\mu_m f_\text{A}(\V{z}_{\text{A},n,l}^{(j)}|\bm{x}_n,  \bm{X}_n)}{\mu_c f_c(\V{z}_{\text{A},n,l}^{(j)})}, &\text{$a_{\text{A},n,l}^{(j)} = 1$}\\ 
		1, & \text{$a_{\text{A},n,l}^{(j)} = 0$}
	\end{cases}
	\label{equ: active pseudo-measurement likelihood}
\end{align}
and
\begin{align}
	&{g}_{\text{P}}(\V{z}_{\text{P},n,l}^{(j,j')}|\bm{x}_n, \bm{X}_n, a_{\text{P},n,l}^{(j,j')})\nn\\
	&\hspace*{15mm}= \begin{cases}
		\frac{\mu_m f_\text{P}(\V{z}_{\text{P},n,l}^{(j,j')}|\bm{x}_n,  \bm{X}_n)}{\mu_c f_c(\V{z}_{\text{P},n,l}^{(j,j')})}, &\text{$a_{\text{P},n,l}^{(j,j')} = 1$}\\ 
		1, & \text{$a_{\text{P},n,l}^{(j,j')} = 0$}
	\end{cases}
	\label{equ: passive pseudo-measurement likelihood}
\end{align}
respectively.

\subsection{Joint Posterior and Factor Graph}
\begin{figure}[t]
	\centering
	\includegraphics{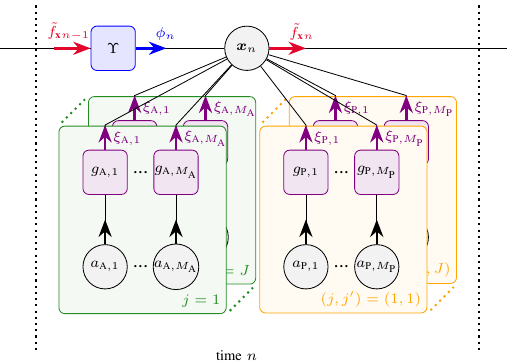}
	\caption{Factor graph representing the factorization of the joint posterior \ac{pdf} in \eqref{equ: joint posterior} and the messages according to the SPA.
	For following short notations are used: $M_{\text{A}} \triangleq M_{\text{A},n}^{(j)}$, $M_{\text{P}} \triangleq M_{\text{P},n}^{(j)}$, $a_{\text{A},l} \triangleq a_{\text{A},n,l}^{(j)}$, $a_{\text{P},l} \triangleq a_{\text{P},n,l}^{(j,j')}$, ${g}_{\text{A},l} \triangleq {g}_{\text{A},n,l}^{(j)}$, ${g}_{\text{P},l} \triangleq {g}_{\text{P},n,l}^{(j,j')}$, $\xi_{\text{A},l}\triangleq \xi_{\text{A},n,l}^{(j)}$, $\xi_{\text{P},l}\triangleq \xi_{\text{P},n,l}^{(j,j')}$.}
	\label{fig:factorGraph}
	\vspace{-6mm}
\end{figure}
We model the evolution of $\bm{x}_n$ over time $n$ as an independent first-order Markov process, which is defined by the state transition PDF $f(\bm{x}_n|\bm{x}_{n-1})$.
We assume that the measurements $\V{z}_{\text{A},n}$ and $\V{z}_{\text{P},n}$ are observed and thus fixed. 
Let $\bm{x} = [\bm{x}_1 ^\text{T}\; \cdots  \bm{x}_n^\text{T}]^\text{T}$, $\bm{a}_{\text{A}}= [\bm{a}_{\text{A},1} ^\text{T}\; \cdots  \bm{a}_{\text{A},n}^\text{T}]^\text{T}$, $\bm{a}_{\text{P}}= [\bm{a}_{\text{P},1} ^\text{T}\; \cdots  \bm{a}_{\text{P},n}^\text{T}]^\text{T}$, $\bm{z}_{\text{A}}= [\bm{z}_{\text{A},1} ^\text{T}\; \cdots  \bm{z}_{\text{A},n}^\text{T}]^\text{T}$, and $\bm{z}_{\text{P}}= [\bm{z}_{\text{P},1} ^\text{T}\; \cdots  \bm{z}_{\text{P},n}^\text{T}]^\text{T}$.
According to the Bayes's rule and other related  independence assumptions\cite{MeyerProc2018}, the joint posterior PDF of all estimated states for time $n$ and all $J$ anchors can be derived as
\begin{align}
	&f(\bm{x},\bm{a}_{\text{A}}, \bm{a}_{\text{P}} | \V{z}_{\text{A}}, \V{z}_{\text{P}})\nn\\ 
	&\hspace{6mm}\propto f(\bm{x}_0) \prod_{n'=1}^{n} \Upsilon(\bm{x}_{n'}|\bm{x}_{n'-1}) \nn\\
	&\hspace{10mm} \times \prod_{j=1}^{J} \prod_{l=1}^{M_{\text{A},n'}^{(j)}} {g}_{\text{A}}(\bm{z}_{\text{A},n',l}^{(j)}|\bm{x}_{n'},\bm{X}_{n'}, a_{\text{A},n',l}^{(j)})\nn\\
	&\hspace{10mm} \times \prod_{j'=1}^{J} \prod_{l=1}^{M_{\text{P},n'}^{(j,j')}} {g}_{\text{P}}(\bm{z}_{\text{P},n',l}^{(j,j')}|\bm{x}_{n'},  \bm{X}_{n'}, a_{\text{P},n',l}^{(j,j')})
	\label{equ: joint posterior}
\end{align}
where $\Upsilon(\bm{x}_n|\bm{x}_{n-1}) \,\triangleq f(\bm{x}_n|\bm{x}_{n-1})$ is the state transition function.
Fig.~\ref{fig:factorGraph} is the factor graph that represents the factorization of the joint posterior in (\ref{equ: joint posterior}).

\subsection{Problem Formulation and Solution}
Our aim is to estimate the agent state ${\RV{x}}_n$ in a sequential way based on the Bayesian framework.  
The agent state is estimated by calculating the \ac{mmse} \cite{Kay1993} of each time step
\begin{equation}\label{eq:mmse}
\hat{\bm{x}}^\text{MMSE}_{n} \,\triangleq \int \rmv \bm{x}_{n} \, f(\bm{x}_{n} | \bm{z} )\, \mathrm{d}\bm{x}_{n} \,
\end{equation}
with $\hat{\bm{x}}^{\text{MMSE}}_n = [\hat{\bm{p}}^{\text{MMSE T}}_n \, \hat{\bm{v}}^{\text{MMSE T}}_n \,\hat{\bm{b}}^{\text{MMSE T}}_n]^\text{T}$. 

The state estimate in (\ref{eq:mmse}) is obtained by calculating the marginal posterior PDFs from the joint posterior PDFs by applying the \ac{spa}\cite{KschischangTIT2001}.
Due to the complexity of the integrals used in the SPA message calculations, analytical solutions are not feasible. As a result, we employ a particle-based representation of the message according to\cite{VenusTWC2023}. This approach allows us to approximately compute the required probabilities while maintaining computational efficiency.


\section{Results}\label{sec:results}


\subsection{Simulation Setup} \label{sec:simulationSetting}

\begin{figure}[t]
	
	\centering
	\setlength{\abovecaptionskip}{0pt}
	\setlength{\belowcaptionskip}{0pt}
	
	\setlength{\figurewidth}{0.28\textwidth}
	\setlength{\figureheight}{0.28\textwidth}
	
    \includegraphics{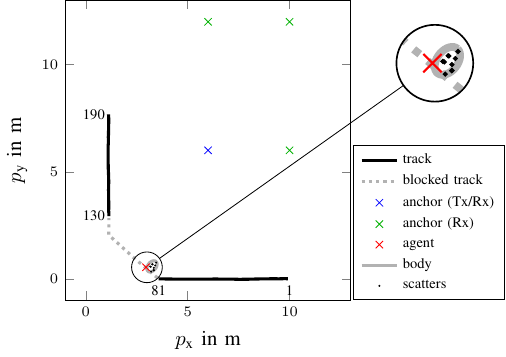}
	\caption{Graphical representation of the synthetic trajectory and generated scatters for passive measurements. The scatters are generated with respect to one anchor at position $(6,6)$ for one snapshot.
	}\label{fig:simulationScenario}
	\vspace{-6mm}
\end{figure}

The proposed algorithm is evaluated using synthetic measurement data according to the scenario presented in Fig.~\ref{fig:simulationScenario} as well as the radio signal models put forward in Section \ref{sec:signal_model}. The agent moves along a trajectory with two distinct direction changes.
The human body, referred to as the extended object, is modeled as an ellipse with the long axis $0.3\,\mathrm{m}$ and short axis $0.2\,\mathrm{m}$. 
The scatters are generated uniformly on the ellipse. 
The bias $[0.25,0.1]^\text{T}\,\mathrm{m}$ between the body center and the mobile agent is set to be constant among the whole trajectory.
The object is observed at $190$ discrete time steps 
at a constant observation rate of $\Delta T = 100\,\mathrm{ms}$.
For active measurements, the signals are transmitted from the mobile agent and  received at $4$ anchors. For passive measurements, the signals are transmitted from one specific anchor at position $(6,6)$ and received at all considered anchors as shown in Fig.~\ref{fig:simulationScenario}.
We assume that the active measurements are blocked by the human body during $[80, 129]$ time steps, which leads to full \ac{olos} situations for part of the track.
The normalized amplitudes are set to $40$ dB at an \ac{los} distance of $1$ m and are assumed to follow free-space path loss, and $\beta_{\vm{q}}=0.2$ and $\beta_{\vm{q}'}=0.8$ are set to make sure the received scattering amplitudes are approximate by $10$ dB lower than the LOS amplitude of each time step. 
The transmitted complex baseband signal $s(t)$ is of root-raised-cosine shape with a roll-off factor of $0.6$ and a bandwidth of $500\,\mathrm{MHz}$.
 
The agent state is represented as $\bm{x}_n = [\tilde{\bm{x}}_n^\text{T}\; \bm{b}_n^\text{T}\;]^\text{T} =[\bm{p}_n^\text{T}\; \bm{v}_n^\text{T}\; \bm{b}_n^\text{T}\;]^\text{T}$.
The PDF of the joint agent state $\bm{x}_n$ is factorized as $f(\bm{x}_n|\bm{x}_{n-1})= f(\tilde{\bm{x}}_n|\tilde{\bm{x}}_{n-1})f(\bm{b}_n|\bm{b}_{n-1})$, where the agent motion, i.e. the state transition \ac{pdf}  $f(\tilde{\bm{x}}_n|\tilde{\bm{x}}_{n-1})$ is described by a linear, constant velocity and stochastic acceleration model\cite[p.~273]{BarShalom2002EstimationTracking}, given as $\tilde{\RV{x}}_n = \bm{A}\, \tilde{\RV{x}}_{n\minus 1} + \bm{B}\, \RV{w}_{n}$. 
The acceleration process $\RV{w}_n$ is i.i.d. across $n$, zero-mean, and Gaussian with covariance matrix ${\sigma_{\text{a}}^2}\, \bm{I}_2$, with ${\sigma_{\text{a}}}$ 
being the acceleration standard deviation, and $\bm{A} \in \mathbb{R}^{\text{4x4}}$ and $\bm{B} \in \mathbb{R}^{\text{4x2}}$ are defined according to \cite[p.~273]{BarShalom2002EstimationTracking}.
The state transition of the bias $\bm{b}_n$, i.e. the state transition PDF  $f(\bm{b}_n|\bm{b}_{n-1})$, is $\bm{b}_n=\bm{b}_{n-1} + \bm{\varepsilon}_{\bm{b}_n}$, where the noise $\bm{\varepsilon}_{\bm{b}_n}$ is i.i.d. across $n$, zeros mean, Gaussian, with covariance matrix ${\sigma_{\bm{\varepsilon}_{\bm{b}}}^2}\, \bm{I}_2$.
The number of particles is set to $I=5000$ for inference during the track, and the particles consist of all considered random variables.
The state-transition variance (STV) are set as ${\sigma_{\text{a}}} = 3\,\mathrm{m/s^2}$ and ${\sigma_{\bm{\varepsilon}_{\bm{b}}}}=0.1\,\mathrm{m}$.
The other simulation parameters are as follows: the detection threshold is $\gamma = 2$ ($6\,\mathrm{dB}$).
The mean number of object-related measurements is set to $\mu_m = 5$, and the scatters are regenerated for each time step and each anchor, for active measurements and passive measurements, respectively.
The mean number of clutter is set to $\mu_c=10$.

\subsection{Performance Evaluation}
\begin{figure}[t]	
	\centering
	\setlength{\abovecaptionskip}{0pt}
	\setlength{\belowcaptionskip}{0pt}
	\includegraphics{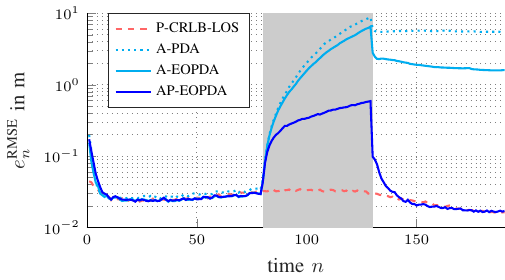}
	\caption{RMSE of the estimated agent position based on numerical simulations. 
	}\label{fig:rmse}
	\vspace{-3mm}
\end{figure}

\begin{figure}[t]
	\centering
	\setlength{\abovecaptionskip}{0pt}
	\setlength{\belowcaptionskip}{0pt}
	\includegraphics{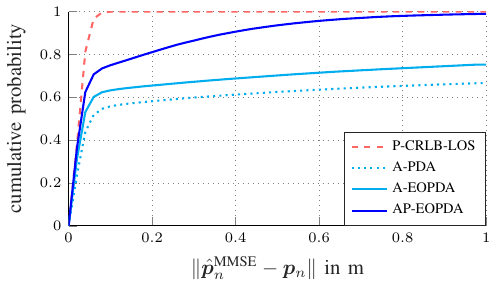}
	\caption{The cumulative distribution of the position errors evaluated by numerical simulations over the whole time period.
	}\label{fig:cdf}
	\vspace{-3mm}
\end{figure}

The numerical results of active-only estimation (``A-PDA" and ``A-EOPDA") as well as joint estimation (``AP-EOPDA") are shown in Fig.~\ref{fig:rmse} and Fig.~\ref{fig:cdf}, respectively.
The \ac{pcrlb} is provided as a performance baseline considering the dynamic model of the agent state \cite{Tichavsky1998, VenusTWC2023}. The ``P-CRLB-LOS" is calculated by assuming the LOS component to all anchors is always available along the whole trajectory, providing a lower bound for the estimation of the object. 
Notably, for both active and passive simulations, we always include the scattering components in the simulated signals, but we compare the results of considering the scattering components in the active case for the data association or not. 
More specifically, ``A-PDA" only considers one object-related measurement, the LOS component, in the PDA, while ``A-EOPDA" and ``AP-EOPDA" take into account multiple object-related measurements, including both the LOS component as well as scatter components.
For joint estimation, we make use of scattering paths from the passive measurements, and assume the passive paths of all anchors are always available along the whole trajectory.

The results are shown in terms of the \ac{rmse} of the estimated agent position $e_{n}^{\text{RMSE}}~=~\sqrt{\E{\norm{\hat{\bm{p}}^{\text{MMSE}}_n -\bm{p}_n}{2}}}$,
and evaluated by numerical simulation with $500$ realizations. 
The RMSE is plotted versus the discrete measurement time $n$ in Fig.~\ref{fig:rmse}. 
Moreover, the cumulative probability of the position errors ${\|{\hat{\bm{p}}^{\text{MMSE}}_n -\bm{p}_n}\|}$ is evaluated by numerical simulations over the whole time period and shown in Fig.~\ref{fig:cdf}.
It can be observed that the RMSE of the joint estimation significantly outperforms the active-only estimation results during and after the full OLOS time steps. 
The uncertainty from the clutter becomes unacceptable for the active estimation during $50$ blocked steps. This leads to outlier occurrence in the active-only results, resulting in a significant difference between the joint and active-only results after the blocked steps.
Besides, comparing the estimation results of ``A-PDA" and ``A-EOPDA", we can find that involving more object-related measurements reduces the number of outliers for the active estimation.
Furthermore, our proposed joint estimation algorithm for the extended object, referred to as ``AP-EOPDA", achieves the ``P-CRLB-LOS" exactly before the OLOS steps, and converges rapidly to the ``P-CRLB-LOS" after the OLOS steps with the aid of passive measurement data. These results demonstrate that the joint estimation algorithm exhibit enhanced robustness and reliability.

\section{Conclusion and Future Work}\label{sec:conclusion}

The key research problem addressed in this paper is how to achieve robust and reliable positioning of a mobile agent (e.g. a mobile device carried by a person) in situations where the direct LOS between the agent and anchors is obstructed by the human body. The human body is modeled as an extended object.
We developed a joint estimation method by applying the \ac{pda} for  both active and passive measurements.
We also consider multiple object-related measurements in the PDA for each time step.
Results show that the joint estimation method achieves significantly lower \ac{rmse}  during and after the \ac{olos} steps compared to the method using only active measurements.
It demonstrates that additional scattering information from passive measurements can help with the positioning during the full \ac{olos} case, and effectively reduce the outliers after the blocked steps. 
Besides, the joint estimation result attains the posterior Cramer-Rao lower bound of LOS component exactly before and after the blocked steps. 
The results illustrate the potential of exploiting scattering paths from the human body for robust localization when line-of-sight signals are unavailable.

According to experiments, scatter points are more prone to occur on the surface of the human body in a specific direction related to the anchors. Thus, our next step is to deal with this physics-based measurement models, and estimate the time-varying extent state together with the kinematic state.
Additionally, our following research can also focus on incorporating real measurement data for the validation of the proposed ``AP-EOPDA", and explore the runtime of our proposed algorithm in a realistic setup.


%

 
 \acrodef{mimo}[MIMO]{multiple input multiple output}
 \acrodef{awgn}[AWGN]{additive white Gaussian noise}
 \acrodef{bw}[BW]{bandwidth}
 \acrodef{blt}[BLT]{bluetooth}
 \acrodef{cdf}[CDF]{cumulative distribution function}
 \acrodef{crlb}[CRLB]{Cram\'er-Rao lower bound}
 \acrodef{dmc}[DMC]{dense multipath component}
 \acrodef{dut}[DUT]{device under test}
 \acrodef{eirp}[EIRP]{equivalent isotropic radiated power}
 \acrodefplural{esl}[ESLs]{electronic shelf labels} 
 \acrodef{los}[LOS]{line-of-sight}
 \acrodef{mf}[MF]{matched filter}
 \acrodef{ml}[ML]{maximum likelihood}
 \acrodef{mpc}[MPC]{multipath component}
 \acrodef{nlos}[NLOS]{non-\ac{los}}
 \acrodef{eot}[EOT]{extended object tracking}
 \acrodef{pcb}[PCB]{printed circuit board}
 \acrodef{pdf}[PDF]{probability density function}
 \acrodef{reb}[REB]{ranging error bound}
 \acrodef{rss}[RSS]{received signal strength}
 \acrodef{smc}[SMC]{specular multipath component}
 \acrodef{snr}[SNR]{signal-to-noise-ratio}
 \acrodef{sinr}[SINR]{signal-to-interference-plus-noise-ratio}
 \acrodef{tdoa}[TDOA]{time difference of arrival}
 \acrodef{tka}[TKA]{trusted keyless access}
 \acrodef{toa}[TOA]{time-of-arrival}
 \acrodef{aoa}[AOA]{angle-of-arrival}
 \acrodef{uwb}[UWB]{ultra wide band}
 \acrodef{mie}[MIE]{method of interval estimation}
 \acrodef{mc}[MC]{Monte Carlo}
 \acrodef{mse}[MSE]{mean squared error}
 \acrodef{ci}[CI]{confidence interval}
 \acrodef{cl}[CL]{confidence level}
 \acrodef{pdp}[PDP]{power delay profile}
 \acrodef{dps}[DPS]{delay power spectrum}
 \acrodef{dm}[DM]{dense multipath}
 \acrodef{nlike}[NLIKE]{normalized likelihood}
 \acrodef{zzb}[ZZB]{Ziv-Zakai bound}
 \acrodef{ut}[UT]{unscented transform}
 \acrodef{glrt}[GLRT]{generalized likelihood ratio test}
 \acrodef{mse}[MSE]{mean squared error}
 \acrodef{rmse}[RMSE]{root mean squared error}
 \acrodef{nnlike}[NNLIKE]{normalized noise-free likelihood}
 \acrodef{stdv}[STDV]{standard deviation}
 \acrodef{rv}[RV]{random variable}
 \acrodef{bp}[BP]{belief propagation}
 \acrodef{pda}[PDA]{probabilistic data association}
 \acrodef{mp}[MP]{multipath}
 \acrodef{pmf}[PMF]{probability mass function}
 \acrodef{pdaf}[PDAF]{probabilistic data association filter}
 \acrodef{pdaai}[AIPDA]{amplitude-information \ac{pda}}
 \acrodef{olos}[OLOS]{obstructed line-of-sight}
 \acrodef{spa}[SPA]{sum-product algorithm}
 \acrodef{mmse}[MMSE]{minimum mean-square error}
 \acrodef{lhf}[LHF]{likelihood function}
 \acrodef{fa}[FA]{false alarm}
 \acrodef{ceda}[CEDA]{channel estimation and detection algorithm} 
 \acrodef{pcrlb}[P-CRLB]{posterior Cram\'er-Rao lower bound}
 \acrodef{slam}[SLAM]{simultaneous localization and mapping}
 \acrodef{mpslam}[MP-SLAM]{multipath-based SLAM}
 \acrodef{va}[VA]{virtual anchor}
 \acrodef{dnr}[DNR]{dense-to-noise ratio}
 \acrodef{aednn}[AE-DNN]{auto encoder deep neural network}   
 \acrodef{gpr}[GPR]{gaussian process regression}  
 \acrodef{ae}[AE]{auto encoder}



\bibliographystyle{IEEEtran}
\bibliography{IEEEabrv,joint_active_passive_tracking,EOT_and_PDA,signalModelingEA,references}

\end{document}